\documentclass[sigconf]{acmart}
\usepackage{enumitem}
\usepackage{hyperref}

\AtBeginDocument{%
  }

\setcopyright{none}
\renewcommand\footnotetextcopyrightpermission[1]{}

\begin{document}

\title{UIBenchKit: A unified toolkit for design-to-code model evaluation}


\author{Chinh T. Le}
\affiliation{%
  \institution{Singapore Management University}
  \country{Singapore}
}
\email{tc.le.2025@mitb.smu.edu.sg}

\author{Trevor Ong Yee Siang}
\affiliation{%
  \institution{Singapore Management University}
  \country{Singapore}
}
\email{trevor.ong.2024@computing.smu.edu.sg}

\author{Jingyu Xiao}
\affiliation{%
  \institution{The Chinese University of Hong Kong}
  \country{China}
}
\email{jyxiao@link.cuhk.edu.hk}

\author{Yuxuan Wan}
\affiliation{%
  \institution{The Chinese University of Hong Kong}
  \country{China}}
\email{yxwan@link.cuhk.edu.hk}

\author{Yintong Huo}
\authornote{Yintong Huo is the corresponding author}
\affiliation{%
  \institution{Singapore Management University}
  \country{Singapore}
}
\email{ythuo@smu.edu.sg}


\begin{abstract}
Recent years have seen substantial progress in automated design-to-code generation, with many methods proposed for generating HTML and CSS from webpage screenshots. However, the absence of a standardized evaluation platform makes it difficult to compare these methods fairly, limiting both practical adoption and systematic research progress. To bridge this gap, we introduce UIBenchKit, an open-source, integrated toolkit designed to unify the evaluation of design-to-code tasks. 
UIBenchKit abstracts the complexities of environment setup, model inference, and code rendering, offering researchers a plug-and-play architecture to compare various methods under consistent settings. In addition, it offers an analytical interface for comparison across multiple metrics. Using UIBenchKit, we conduct a benchmarking study of existing tools and derive several findings that highlight directions for future improvement.
By providing a streamlined environment for both experimentation and evaluation, UIBenchKit aims to accelerate future benchmarking and innovations in web engineering. 
The evaluation platform and toolkit are available at the project page \url{https://www.uibenchkit.com/}. 
\end{abstract}


\maketitle

\section{Introduction}
The translation of visual user interface (UI) mockups into functional front-end implementations (referred to as \textit{UI2Code} or \textit{design-to-code}), sits at the intersection of design and web development. This step remains challenging and costly, as it demands visual fidelity and correct implementation while coordinating designers and developers. 

Recently, Multimodal Large Language Models (MLLMs) have demonstrated remarkable capabilities in directly generating structured HTML and CSS from visual inputs, catalyzing a wave of both research and commercial interest in this area.
In the commercial space, dedicated UI2Code products including Vercel's v0, Lovable, Bolt.new, and Figma, have collectively attracted millions of active users \cite{saastr_v0}. On the research front, a growing body of work has emerged on UI2Code, spanning methods that improve generation quality  \cite{Wan_2025, Gui_2025, gui2025uicopilotautomatinguisynthesis, Wu_2025}, benchmarks that evaluate model performance \cite{si2025design2codebenchmarkingmultimodalcode}, and recent efforts that extend UI2Code generation from static UI mockups to interactive webpages \cite{xiao2025interaction2code}.

Despite this rapid progress, there is a lack of a standardized evaluation toolkit for UI2Code research. Individual evaluations are tightly coupled to specific LLM backbones, configurations, and prompts \cite{si2025design2codebenchmarkingmultimodalcode, Wan_2025, Wu_2025}. 
Moreover, there is also no unified evaluation metric.
Without a standardized pipeline, it is unclear whether reported improvements reflect stronger methods or simply differences in prompts or evaluation protocols. As a result, this makes it difficult for practitioners and researchers to choose and compare tools.

We believe that a reliable evaluation platform for UI2Code must satisfy several key requirements. First, it must be \textit{modularized} to isolate the effects of different factors, thereby enabling controlled and reproducible experiments. Second, it has to be \textit{extensible} to incorporate new benchmarks and methods. 
Third, given the complexity of UI2Code evaluation, which spans visual, structural, and textual evaluation metrics \cite{si2025design2codebenchmarkingmultimodalcode, Wu_2025}, the platform should provide a \textit{data analytical interface} to help practitioners compare tools.
\begin{figure*}[t]
    \centering
    \includegraphics[width=\textwidth]{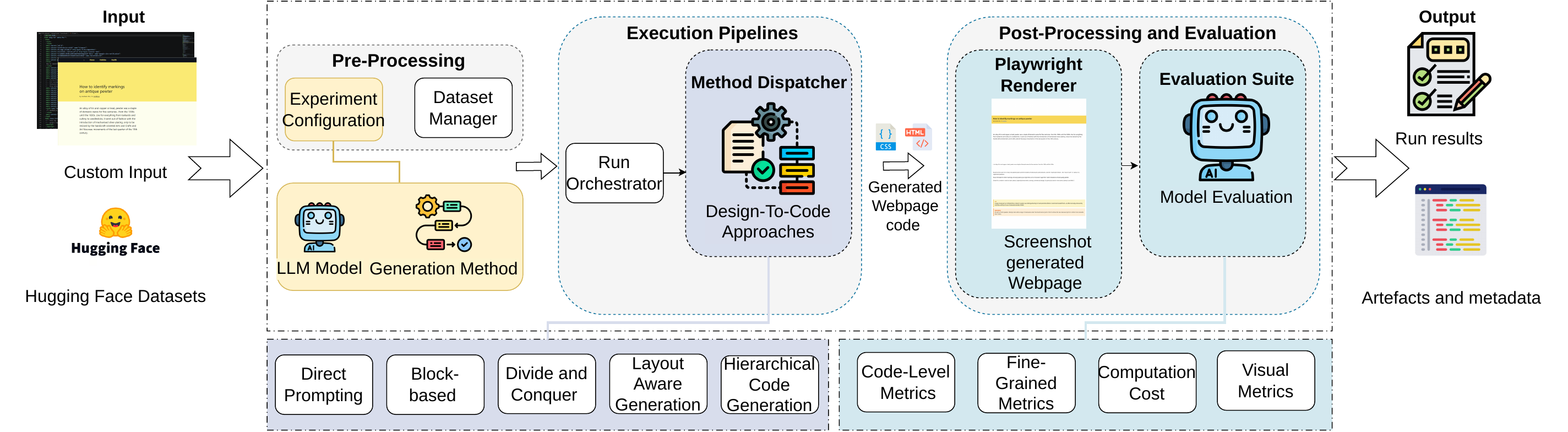}
    \Description{A block diagram illustrating the system architecture, showing the flow of data from the input module through the processing layers to the final output generation.}
    \caption{UIBenchKit System Architecture Design}
    \label{fig:full_width_diagram}
\end{figure*}

To address these needs, we build \textbf{UIBenchKit}, a unified and extensible toolkit for developing, reproducing, and benchmarking design-to-code methods. UIBenchKit abstracts the end-to-end workflow into modular components. It encompasses integration with various MLLMs, method execution, and evaluation. This allows users to test and compare multiple methodologies through a simple interface.
We envision that UIBenchKit enables more transparent and reproducible evaluation of UI2Code frameworks in future research.
In summary, UIBenchKit makes three main contributions.

\begin{enumerate}[leftmargin=*]
    \item \textbf{A unified execution framework for UI2Code research.} UIBenchKit provides a standardized pipeline for evaluating design-to-code tools across datasets, backbone LLMs, and methodologies. It handles dataset preparation, method execution, and output generation through a unified interface.
    
    \item \textbf{An analytical interface for multi-dimensional comparison.} UIBenchKit provides a reporting and visualization interface that helps users inspect generated outputs and compare methods across multiple dimensions. 
    
    \item \textbf{A large-scale benchmarking study.} Using UIBenchKit, we benchmark 16 models, 5 methodologies, and 2 datasets, which total to 832 instances under a unified setting, yielding a comprehensive view of current performance in terms of visual fidelity, structural accuracy, and computational overhead. 
\end{enumerate}


\section{The Design of UIBenchKit}

\subsection{System Architecture}

Figure 1 illustrates UIBenchKit's end-to-end evaluation system. The evaluation begins with the \textit{Dataset Manager}, which prepares benchmark datasets or custom screenshot folders into a shared input format. 
This ensures that all methods operate on the same input structure and preprocessing conventions. 
After the inputs are normalized, the \textit{Run Orchestrator} records the experiment configuration, tracks the progress for each instance, and stores the generation outputs and metadata. 
This provides reproducibility and supports recovery for long-running or failure-prone benchmark runs.

During execution, the \textit{Method Dispatcher} route prepared inputs to various UI2Code models.
Although the methods differ in decomposition, generation, and assembly strategies, UIBenchKit exposes them through a common execution interface. 

For each LLM call made by a selected method, UIBenchKit abstracts provider-specific inference details such as input formatting, API calls, backend routing, and token accounting. 
This allows a method to be easily evaluated across different MLLM families. 
Once the method produces HTML, the \textit{Post-processing and Evaluation} layer renders the generated code into screenshots using Playwright and applies an extensible set of evaluation metrics.

\subsection{Tool Implementation}
Our implementation of UIBenchKit consists of two main modules:
\begin{itemize}[leftmargin=*]
\item \textbf{Backend API}: The core execution engine is implemented in Python using Flask \cite{flask_docs}. It serves as the entry point for run submission, status tracking, and artifact access.
\item \textbf{Web Frontend}: For interactive use, UIBenchKit provides a browser-based frontend implemented as a Single Page Application using React \cite{react_docs}, TypeScript \cite{typescript_docs}, and Tailwind CSS \cite{tailwind_docs}. This interface allows users to upload screenshots or benchmark-style folders, select models and methods, inspect generated outputs, and review evaluation results.
\end{itemize}
\subsection{Reproducing Benchmark Runs}
UIBenchKit supports reproducible experiments through a shared CLI and REST API workflow. Users specify an input source, target model, and generation method, then submit the run through a unified interface. During execution, UIBenchKit exposes status polling and stores intermediate artifacts, allowing users to monitor progress and recover incomplete runs. Upon completion, it returns a consolidated report containing configurations, generated HTML, rendered screenshots, token statistics, and evaluation scores.

\subsection{Data Analytical Interface}
UIBenchKit provides an analytical reporting module for interpreting evaluation results across models and methods. Upon run completion, the toolkit aggregates data into a unified report supporting multi-dimensional comparisons across models and methods. Users can compare model-method combinations across visual similarity, structural accuracy, fine-grained UI matching, and computational overhead. Crucially, the interface links quantitative metrics directly to underlying artifacts, such as input screenshots, generated HTML, and rendered webpages. 
This traceability allows users to determine not just \textit{which} method excels, but \textit{why} specific failures occur. The interface module complements the execution framework and delivers actionable insights for researchers and practitioners.

\begin{figure}[h]
    \centering
    \includegraphics[width=\linewidth]{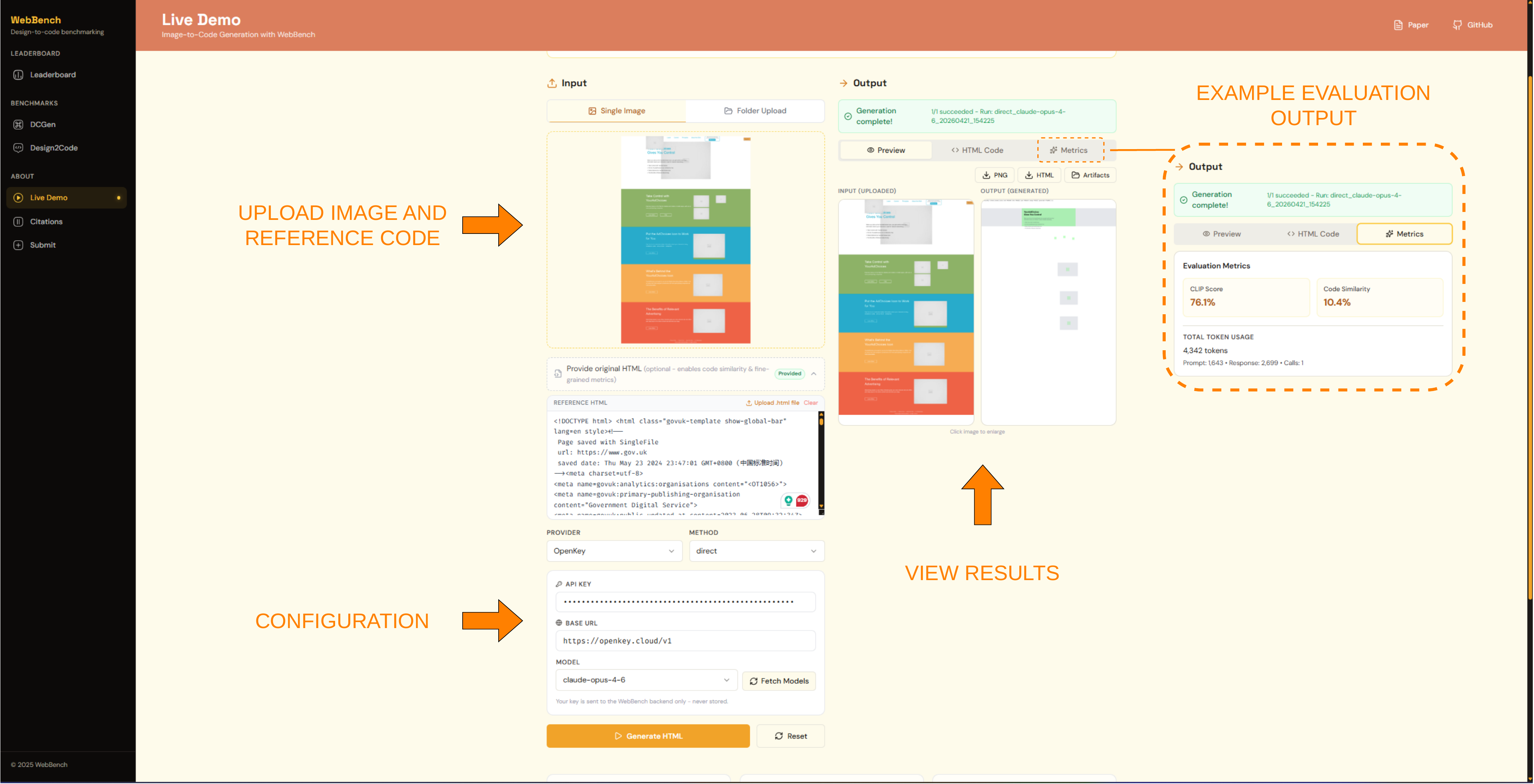}
    \caption{UIBenchKit Graphical User Interface}
    \label{fig:full_width_diagram}
    \Description{UIBenchKit Graphical User Interface}
\end{figure}

\section{User Demo of UIBenchKit}
To complement the CLI and API workflows for large-scale experiments, UIBenchKit provides a lightweight web interface tailored for quick exploratory analysis and qualitative inspection. This GUI facilitates an interactive demo on targeted UI examples through following workflow:
\begin{enumerate}[leftmargin=*]
    \item \textbf{Upload UI examples.}
    The user uploads one or more target UI screenshots.
    \item \textbf{Select comparison settings.}
    The user selects specific methods and target LLMs for comparison.

    \item \textbf{Run generation.}
    UIBenchKit processes the selected combinations through the same backend generation and rendering pipelines as the CLI and API.

    \item \textbf{Inspect results.}
    The interface presents the input screenshot alongside the generated HTML, rendered output, and evaluation metrics. This enables a rapid case-to-case comparison.
    
\end{enumerate}
\section{Evaluation}
To demonstrate the practicality of UIBenchKit, we utilized the framework to conduct a large-scale empirical study across diverse models and methodologies. This study includes 16 leading LLMs and 5 open-source methodologies on two popular datasets.
The evaluation was orchestrated using the tool's API entry point, ensuring standardized preprocessing and identical rendering environments across all test runs. 
\subsection{Experimental Setup}
\begin{enumerate}[leftmargin=*]
    \item \textbf{Datasets:} 
    We evaluated the generated frontend code on two widely recognized benchmarks: Design2Code \cite{si2025design2codebenchmarkingmultimodalcode} (484 real-world webpage mockups) and DCGen \cite{Wan_2025} (348 distinct webpage designs).

    \item \textbf{Models:} 
    Rather than limiting our evaluation to a single architecture, we benchmark 16 distinct Multimodal Large Language Models. These span state-of-the-art proprietary models (e.g., GPT, Claude, and Gemini families) and open-source ones (e.g., Qwen and LLaMA families).

    \item \textbf{Methods:} 
    We evaluate five representative UI2Code methodologies supported by UIBenchKit. We reproduce the open-source methodologies using their original implementations.
    \begin{itemize}[leftmargin=*]
        \item \textit{Direct prompting} We evaluate the basic design-to-code capability of each LLM by directly prompting them. The unified prompt is: \textit{"Here is a prototype image of a webpage. Return a single piece of HTML and Tailwind CSS code to reproduce exactly the website. Use "placeholder.png" to replace the images. Pay attention to things like size, text, position, and color of all the elements, as well as the overall layout. Respond with the content of the HTML+Tailwind CSS code."};
        \item \textit{DCGen} \cite{Wan_2025}, a divide-and-conquer approach that segments the input UI into regions, generates region-level code, assembles them into a complete layout, and refines the result through iterative rendering and candidate selection;
        \item \textit{LaTCoder} \cite{Gui_2025}, a Layout-as-Thought method that decomposes a UI into layout-aware blocks, generates code for each block using CoT-based prompts, and assembles the final page with dynamic selection;
        \item \textit{LayoutCoder} \cite{Wu_2025}, a layout-guided method that builds a UI layout tree from element relations and recursive block projection, then fuses MLLM-generated atomic-region code snippets into the final webpage;
        \item \textit{UICopilot} \cite{gui2025uicopilotautomatinguisynthesis}, a hierarchical method that predicts a coarse DOM tree with bounding boxes, generates HTML/CSS for leaf-node regions, and refines global styles while preserving the structure.
    \end{itemize}

\end{enumerate}
\subsection{Evaluation Metrics}
For our experimental setup, we utilize screenshots of live webpages to simulate the input design mockups $I_o$, relying on their underlying HTML and CSS as the ground-truth code $C_o$. We measure generation performance across three primary dimensions: high-level similarity, fine-grained element matching, and resource efficiency.
\begin{itemize}[leftmargin=*]
\item \textbf{High-Level Similarity:} We measure the overall closeness of the outputs through two metrics: \textit{Structural Code Similarity}, utilizing Normalized Levenshtein Distance (S = 1 - distance/($l_1$ + $l_2$)) \cite{rapidfuzz2024} between $C_g$ and $C_0$, and \textit{Visual Image Similarity} via CLIP Score \cite{si2025design2codebenchmarkingmultimodalcode}, employing CLIP-ViT-B/32 embeddings to score the resemblance of $I_g$ to $I_o$.
\item \textbf{Fine-Grained Element Matching:} To capture detailed performance nuances missed by macro-level evaluations, we adopt the element-matching suite from Si et al \cite{si2025design2codebenchmarkingmultimodalcode}. This approach detects and aligns visual blocks between $I_0$ and $I_g$, assessing accuracy across four specific dimensions: \textit{Block-match} (ratio of matched to total block sizes), \textit{Text similarity} (Sørensen-Dice coefficient), \textit{Color similarity} (CIEDE2000 formula), and \textit{Position similarity} (block center alignment).
\item \textbf{Resource Consumption:} Because inference cost has emerged as a primary bottleneck limiting the practical scalability and commercial adoption of LLMs \cite{zhuang2025benchmarkseconomicsaiinference}, we systematically track and evaluate the \textit{token consumption} for each method. Specifically, we calculate the average number of text and vision tokens consumed by the respective MLLMs per generation, providing a transparent benchmark for computational overhead.
\end{itemize}

\subsection{Evaluation Results}
\begin{figure}[t]
    \centering
    \includegraphics[width=\linewidth]{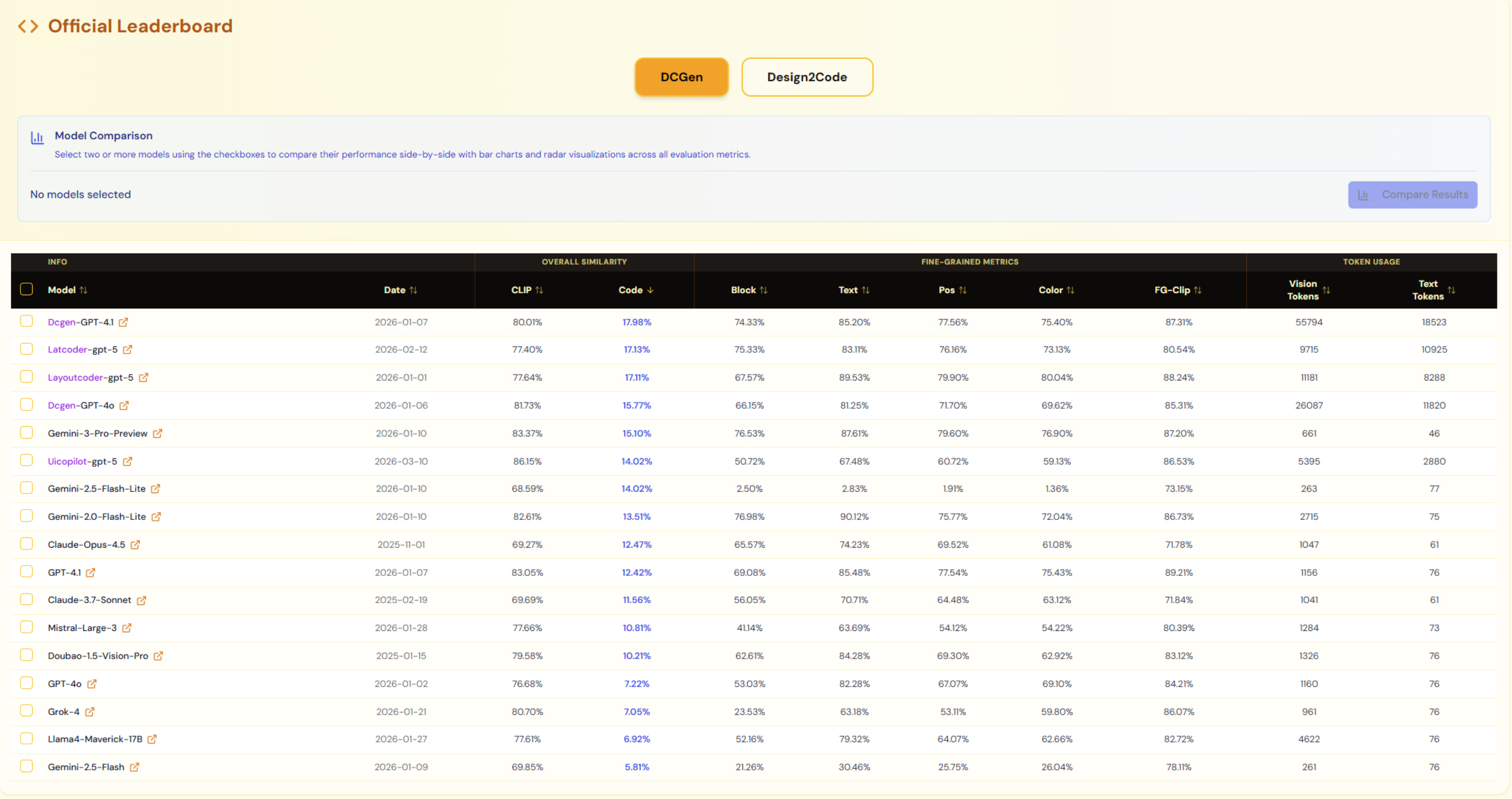}
    \caption{Evaluation result available on the project website}
    \label{fig:full_width_diagram}
    \Description{Evaluation result}
\end{figure}
Due to space constraints, the full evaluation results are available on the project website at \url{https://www.uibenchkit.com/}. Figure 3 illustrates the leaderboard view used to compare benchmark results across models and methods. Our experimental findings show that decomposition-based UI-to-code frameworks can improve layout fidelity on more complex webpages, but their gains are not uniform across all evaluation metrics. For example, LaTCoder performs strongly on structural metrics, particularly in foreground block matching and positioning, whereas direct prompting remains competitive on simpler cases. We also observe that global similarity metrics, such as CLIP, can obscure layout errors, so fine-grained-level metrics provide a clearer picture of practical UI fidelity.

\section{Lesson Learnt}
While running our standardized evaluation pipeline, we observed that many UI2Code methods, despite their distinct designs, share a common segmentation-generation pattern: they first decompose screenshots into visual or structural units before generating and assembling code. To better understand recurring failures, we analyzed cases where the generated webpage was incomplete or invalid, leading to two key insights.
\begin{itemize}[leftmargin=*]
\item \textbf{Optimization:} While decomposition-based pipelines improve scalability, they shift the problem from generation complexity to intermediate representation accuracy, making overall performance highly sensitive to early-stage errors. Its benefits include a reduced context length for LLM-based code generation, which simplifies LLM reasoning of local UI elements. The modularity of UI elements also enables parallel code generation, which reduces generation time. However, our experiments also showed that this optimization introduces a systemic trade-off. Errors in early stages, such as inaccurately identifying bounding boxes are amplified downstream. The framework either fails when reaching the threshold number of bounding boxes allowed or it wastefully generates redundant code chunks. 
\item \textbf{Image segmentation:} Image segmentation is the primary source of failure, as current methods rely heavily on low-level visual cues that do not reliably capture the semantic structure of modern UIs. We have come to realize that there is no "one size fits all" image segmentation algorithm. Approaches like the one proposed in LayoutCoder \cite{Wu_2025} assumes that elements are xy-axis aligned. This means that the algorithm struggles to segment angled elements accurately. LatCoder \cite{Gui_2025} on the other hand relies heavily on identifying strong visual boundaries, leading to decorative lines, borders, and even text at times being over-segmented. The issues identified here stems from the fact that images do not encode the visual hierarchy that is inherently obvious to us humans. 
\end{itemize}
Our research has found that LLM code generation capabilities have improved tremendously, aided by frameworks which segments images into smaller visual blocks. However, the most significant hurdle in UI2Code generation remains the limitations of existing image segmentation techniques. Future image segmentation systems should attempt to incorporate structure-aware representations of the image segments. In fact, UICopilot's \cite{gui2025uicopilotautomatinguisynthesis} assembly stage makes a good attempt at incorporating both structural references and semantic relationships between UI elements. UICopilot \cite{gui2025uicopilotautomatinguisynthesis} not only produces image segments but also uses a Transformer-based decoder to infer the DOM tree of a webpage. We believe that this is a direction that has potential for exploration. We hope that UIBenchKit will continue to provide key insights that will inspire subsequent research efforts in UI2Code methods in the same way we gained insights through our experiments. 



\section {CONCLUSION}
In this paper, we introduce UIBenchKit, an extensible and unified framework designed to address the current fragmentation in automated design-to-code research. We demonstrated the platform's robustness by seamlessly integrating 5 distinct generation methodologies and 16 diverse Multimodal Large Language Models (MLLMs), culminating in a comprehensive empirical benchmark. By providing both a RESTful orchestration API and a high-throughput CLI, UIBenchKit significantly lowers the barrier to entry for reproducing state-of-the-art results and conducting fair, standardized evaluations. In our future works, we will focus on expanding support for dynamic web components and integrating modern frontend frameworks like React to further bridge the gap between multimodal AI research and practical software engineering workflows.


\bibliographystyle{ACM-Reference-Format}
\bibliography{references}









\end{document}